\documentclass[fleqn,usenatbib]{mnras}

\usepackage{txfonts}

\usepackage[T1]{fontenc}

\DeclareRobustCommand{\VAN}[3]{#2}
\let\VANthebibliography\thebibliography
\def\thebibliography{\DeclareRobustCommand{\VAN}[3]{##3}\VANthebibliography}

\usepackage{graphicx}	
\usepackage[dvipsnames]{xcolor} 
\usepackage[normalem]{ulem}

\title[Outflows from 3C\,293]{Spatially resolved observations of outflows in the radio loud AGN of UGC\,8782}

\author[R. A. Riffel et al.]{Rogemar A. Riffel,$^{1}$\thanks{E-mail: rogemar@ufsm.br}
Rog\'erio Riffel,$^{2}$ Marina Bianchin,$^{1,3}$ Thaisa Storchi-Bergmann,$^2$ \newauthor Gabriel Luan Souza de Oliveira,$^{1}$ Nadia L. Zakamska$^{4}$.
\\
$^{1}$Departamento de F\'isica, CCNE, Universidade Federal de Santa Maria, 97105-900 Santa Maria, RS, Brazil
\\
$^{2}$Departamento de Astronomia, IF, Universidade Federal do Rio Grande do Sul, CP 15051, 91501-970, Porto Alegre, RS, Brazil\\ 
$^{3}$Department of Physics and Astronomy, 4129 Frederick Reines Hall, University of California, Irvine, CA 92697, USA\\
$^{4}$Department of Physics \& Astronomy, Johns Hopkins University, Bloomberg Center, 3400 N. Charles St, Baltimore, MD 21218, USA.
}

\date{Accepted XXX. Received YYY; in original form ZZZ}

\pubyear{2022}

\begin{document}
\label{firstpage}
\pagerange{\pageref{firstpage}--\pageref{lastpage}}
\maketitle

\begin{abstract}
We use optical Integral Field Spectroscopy (IFU) to study the gas emission structure and kinematics in the inner 3.4$\times$4.9 kpc$^2$ region of the galaxy UGC\,8782 (3C\,293), host of a radio loud Active Galactic Nucleus (AGN). The observations were performed with the GMOS-IFU on the Gemini North telescope, resulting in a spatial resolution of $\sim725$\,pc at the distance of the galaxy. While the stars present ordered rotation following the orientation of the large scale disc, the gas shows  a disturbed kinematics. The emission-line profiles present two kinematic components: a narrow ($\sigma\lesssim200$ km\,s$^{-1}$) component associated with the gas in the disc of the galaxy and a broad ($\sigma\gtrsim200$ km\,s$^{-1}$) component produced by gas outflows. Emission-line  ratio diagrams indicate that the gas in the disc is excited by the AGN radiation field, while the emission of the outflow includes additional contribution of shock excitation due to the interaction of the radio jet with the environment gas. Deviations from pure rotation, of up to 30\,km\,s$^{-1}$, are observed in the disc component and likely produced by a previous merger event. The broad component is blueshifted by  $\sim150-500$\,km\,s$^{-1}$ relative to the systemic velocity of the galaxy in all locations. We construct radial profiles of the mass outflow rate and kinetic power of the ionized gas outflows, which have the maximum values at $\sim1$\,kpc from the nucleus with peak values of $\dot{M}_{\rm out,\Delta R}=0.5\pm0.1$ M$_\odot$\,yr$^{-1}$ and $\dot{K}_{\rm out,\Delta R} =$(6.8$\pm$1.1)$\times$10$^{41}$ erg\,s$^{-1}$. The kinetic coupling efficiency of these outflows are in the range of 1--3 per cent, indicating that they could be powerful enough to affect the star formation in the host galaxy as predicted by theoretical simulations. 

\end{abstract}




\section{Introduction}
\label{sec:introduction}

Feedback from active galactic nuclei (AGN) is claimed to play a major role in shaping massive galaxies, by quenching star formation and transforming them from star-forming to quiescent galaxies    \citep[e.g.][]{dimatteo05,fabian2012,kormendy13,Harrison_2018}. 
A critical ingredient in the rapid quenching of star formation in AGN hosts are multi-phase gas outflows, which can push gas out of the galaxy or redistribute it, preventing it from collapsing to form stars. However, understanding the acceleration mechanisms of such outflows remains an important unknown piece of the galaxy evolution puzzle. This question can be addressed by spatially resolved observations of tracers of the multiple gas phases of AGN driven winds \citep[e.g.][]{feruglio10,feruglio15,liu13,may17,ramos-almeida17,shimizu19,rupke19,couto20,Oliveira21,cameron21,rogemar06_eso,rogemar20_n1275,rogemar21_cyg,rogemar_2023_extended,ruschel-dutra21,speranza22}.

\begin{figure}
    \centering
    \includegraphics[width=0.49\textwidth]{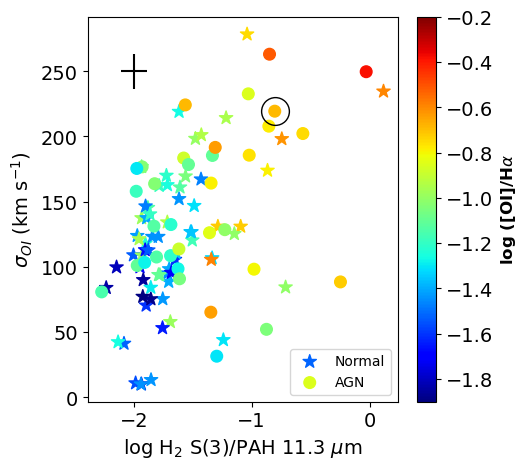} 
\caption{\small [O\,{\sc i}] velocity dispersion vs. the H$_2$S(3)/PAH$\lambda$11.3$\mu$m intensity line ratio for the matched Spitzer-SDSS sample from \citet{rogemar20_spitzer}. The points are color coded according the [O\,{\sc i}]$\lambda$6300/H$\alpha$ values as indicated by the color bar. AGN hosts are shown as filled circles, while normal galaxies are represented by stars. Typical uncertainties are shown in the top-left corner and the open circle represents UGC\,8782.} 
    \label{fig:spitzer}
\end{figure}

Using a sample composed of all galaxies in the entire spectroscopic database of the Spitzer Space Telescope, \citet{lambrides19} found that AGN hosts show a significant excess of H$_2$ emission compared to the expected emission from star formation in normal photo-dissociation regions. In a following up work, \citet{rogemar20_spitzer} cross-correlated the sample of \citet{lambrides19} with the SDSS-III database. One important result of this work is that the excess of H$_2$ emission appears to be strongly related both to emission-line and kinematic shock diagnostics, such as [O{\,\sc i}]$\lambda$6300 flux and velocity dispersion -- as shown in Figure~\ref{fig:spitzer}. These results suggest that the H$_2$ emission excess originates from wind-driven shocks in neutral and molecular gas \citep{hill14}. These conclusions were based on single aperture spectra. Spatially resolved observations of the neutral and ionized gas emission can provide indirect information about the role of AGN winds in the production of the H$_2$ emission excess in the mid-IR.  

It is very difficult to accelerate dense clouds of molecular gas up to velocities high enough to escape galaxies without them being fragmented and the molecules dissociated by the intense radiation field of the AGN. Recent theoretical models suggest that molecules may be formed inside the wind, and their emission may be produced by shock heating of the gas \citep{richings18a,richings18b}.  The CO emission is commonly used as a tracer of cold molecular (T $\sim 100$\,K) outflows in AGN hosts  \citep[e.g.][]{almudena19}, while the hot molecular phase ($T>1000$\,K) can be studied via the H$_2$ near-infrared emission lines \citep[e.g.][]{davies14,fischer17,rogemar21_survey,rogemar22_surveykin,bianchin22}. Not much is known about the warm molecular phase (T $\sim 200$ -- $1000$\,K)  traced by the H$_2$ emission in the the mid-infrared,  which can be studied in a spatially resolved way only now with the JWST. As such observations are very time demanding, it is essential to choose carefully the objects to be studied. 

Here, we present optical integral field unit (IFU) observations of the central region of the galaxy UGC\,8782, which are used to map the neutral and ionized gas emission structure and kinematics. This object was selected as a likely host of strong molecular outflow, by being among the galaxies with the highest values of H$_2$S(3)/PAH$\lambda$11.3$\mu$m, [O{\,\sc i}]$\lambda$6300/H$\alpha$ intensity line ratios and [O{\,\sc i}]$\lambda$6300 velocity dispersion in the sample of \citet{rogemar20_spitzer}. It is identified by a circle in Fig.~\ref{fig:spitzer} and direct observations of the warm molecular hydrogen emission will be obtained as part of an approved cycle 1 JWST proposal (Proposal 1928, PI: Riffel, R. A.).

UGC\,8782 (3C\,293) at a redshift of $z = 0.045$ \citep{sandage66} is a radio loud source \citep{liu02} with nuclear activity classified as a Low-Ionization Nuclear Emission Region \citep[LINER;][]{veron06}. Optical images reveal complex morphology with filamentary dust lanes at hundreds of parsecs and kpc scales, likely as a result of a merger event \citep[e.g.][]{martel99}. This galaxy presents a double-double radio source, with 200 kpc outer lobes oriented approximately along the northwest-southeast direction and  $\sim$4 kpc scale inner lobes along the east-west direction \citep[e.g.][]{machalski16}.  Outflows from the nucleus of UGC\,8782  were observed in neutral hydrogen, with velocities of up to 1400 km\,s$^{-1}$, coming from the western radio hotspot \citep{morganti03,mahony13} and in ionized gas, but with the highest velocities, of up to 1000 km\,s$^{-1}$, co-spatial with the eastern radio knot, at the opposite side of the neutral gas outflows \citep{emonts05,mahony16}. These outflows have been interpreted by the authors as being produced by the interaction of the radio jet with the interstellar medium.  \citet{Kukreti22} combined multi-frequency radio data of 3C\,293, obtained with the International LOw Frequency ARray (LOFAR) telescope (ILT),  Multi-Element Radio Linked Interferometer Network (MERLIN) and Very Large Array (VLA), to map the spectral index over a broad frequency range. They found that the inner radio lobes are part of a young jet-dominated radio source strongly interacting with the interstellar medium, likely driven the ionized outflows mentioned above.
By comparing the spectral indices of the inner radio structure to that from the structures at larger scales, the authors conclude that 3C 293 underwent at least two episodes of nuclear activity.

This paper is organized as follows: Section 2 presents the observations and data reduction procedure, while Section 3 describes the emission line fitting and stellar kinematics measurements. The results are presented in Section 4 and discussed in Section 5, and Section 6 summarizes our conclusions.   We adopt a redshift based distance to UGC\,8782 of 200 Mpc, for which 1 arcsec corresponds to $\sim$970 pc at the galaxy. 
 
 \section{Observations and Data Reduction}
 
 UGC\,8782 was observed on February 23, 2022 with the Gemini-North Multi Object Spectrograph \citep[GMOS,][]{hook04} operating in the Integral Field Unit \citep[IFU,][]{allington-smith02} mode, under the program code GN-2022A-Q-310 (PI: Riffel, R. A.). We used B600-G5323 grating and the one slit mode of the GMOS-IFU, which provides a field of view of 3\farcs5$\times$5\farcs0. The observations of UGC\,8782 were split into six exposures of 1200 sec, half centred at 6250\AA\ and half at 5950\AA, in order to account for the detector gaps. 
 
 The data reduction was performed with the {\sc gemini.gmos} package with the {\sc iraf} software following the standard procedures for spectroscopic data, which includes the subtraction of the bias level, trimming and flat-field correction, background subtraction for each science data, quantum efficiency correction, sky subtraction and wavelength calibration. The {\sc lacos}  algorithm \citep{vandokkum01} was used for cosmic rays removal and finally we performed the flux calibration, using observations of the standard star Feige\,66 to construct the sensitivity function. The final data cubes for each individual exposure have an angular sampling of 0\farcs1$\times$0\farcs1. 
 
 The final data cube for  UGC\,8782 was obtained by median combining the individual exposure data cubes using the peak of the continuum as reference for astrometry corrections among individual cubes. The final datacube covers the spectral region from 4650 to 7500 \AA, which includes the most prominent optical emission lines seen in AGN spectra, from H$\beta$ to [S\,{\sc ii}]$\lambda$6731\AA. The velocity resolution is $\approx$80  km\,s$^{-1}$ as estimated from the full width at half maximum (FWHM) of typical emission lines in the CuAr spectra used in the wavelength calibration and the angular resolution is $\sim$0\farcs75 as estimated from the FWHM field stars in the UGC\,8782 acquisition image, corresponding to the $\sim725$\,pc at the distance of the galaxy.

 \section{Measurements}
 
\subsection{Stellar Kinematics}

We use the Penalized Pixel-Fitting ({\sc ppxf}) method \citep{cappellar04,Cappellari2017,cappellari22} to fit and subtract the stellar component from the observed spectra.   The code finds the best fit of the observed spectra by convolving template spectra with the Line of Sight Velocity Distribution (LOSVD) assumed to be reproduced by Gauss-Hermite series. We fit the whole spectral region, masking out regions of strong emission lines and use as spectral templates those from the MILES-HC library \citep{Westfall_2019}, which has a similar spectral resolution of our GMOS data. We allow {\sc ppxf} to include multiplicative fifth order polynomials to further adjust the continuum shape of the template to the observed spectra.  In addition, we use the {\it clean} parameter of {\sc ppxf} to use the iterative sigma clipping method described in \citet{cappellari02} to reject spectral pixels that deviates more than 3$\sigma$ from the best fit, in order to exclude from the fit unmasked bad pixels, due to remaining sky lines and spurious features.

The output from the {\sc ppxf} code includes measurements of the radial velocity (V$_*$) and stellar velocity dispersion ($\sigma_*$)  at each spaxel, which are used to construct two-dimensional maps presented in Fig. ~\ref{fig:maps}.  In addition, we subtract the best fit model at each spaxel from the observed spectra in order to produce a data cube free of the stellar population component, which is used to measure the emission-line properties.

\subsection{Emission-line Fitting}

 We use the {\sc ifscube} package \citep{daniel_ruschel_dutra_2020_3945237,ruschel-dutra21} to fit the observed emission-line  profiles  by Gaussian curves.  The fit is performed after the subtraction of the stellar component contribution from the observed spectra, and we allow the code to include two Gaussian curves (a $narrow$ and a $broad$ component) per emission line, as indicated by visual inspection of the spectra. 
 We use the {\sc cubefit} routine from the {\sc ifscube} package and fit the following emission lines simultaneously: H$\beta$, [O {\sc iii}] $\lambda\lambda$4959,5007, [O {\sc i}] $\lambda$6300, [N {\sc ii}] $\lambda\lambda$6548,6583, H$\alpha$ and   [S {\sc ii}] $\lambda\lambda$6717,6731. Initial guesses of the Gaussian amplitude, centroid velocity, and velocity dispersion are provided to the code based on measurements of the nuclear spectrum using the {\it splot} {\sc iraf} task, which are used to fit the nuclear spectrum. After a successful fit of the nuclear emission line profiles, the routine fits the surrounding spaxels following a radial spiral loop, using as initial guesses the best fit parameters obtained from successful fits of spaxels at distances smaller than 0\farcs3 from the fitted spaxel, as defined by using the {\it refit} parameter. If the amplitude of one of the fitted Gaussians is smaller than three times the standard deviation of the nearby continuum, only one component is used. To account for possible remaining continuum emission, we also include 4${\rm th}$ order polynomial, which is used to fit the continuum before the fitting of the emission lines.

\begin{figure*}
    \centering
    \includegraphics[width=0.99\textwidth]{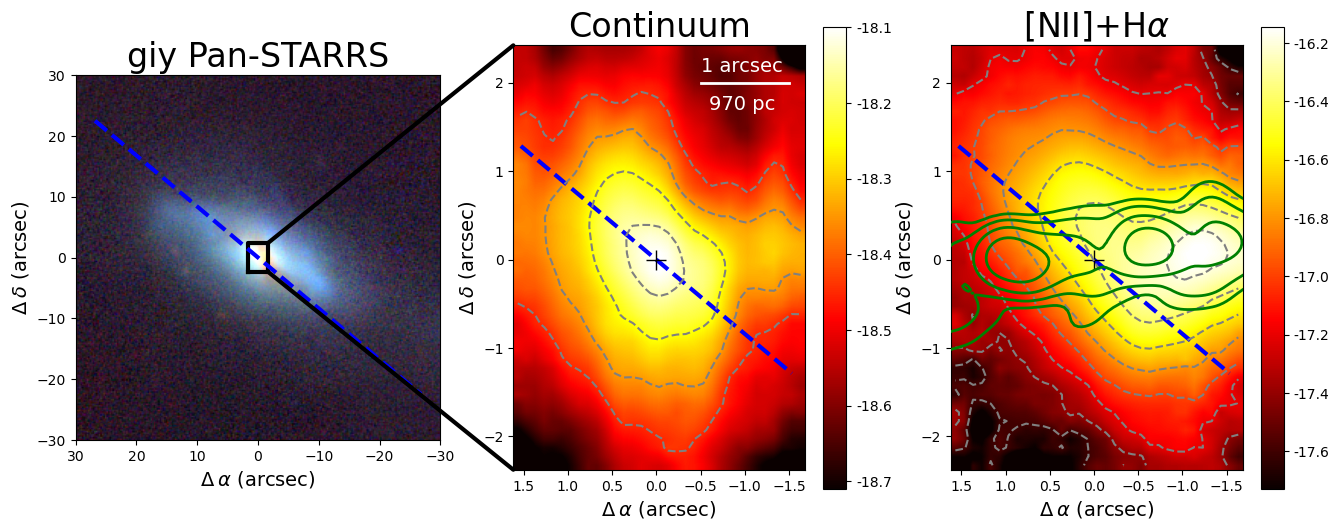} 
\caption{\small Left: Composite image of the 1$^\prime\times$1$^\prime$ {\em g} (4866\AA), {\em i} (4866\AA) and {\em y} (9633\AA) bands of UGC\,8782 from Pan-STARRS data archive \citep{chambers16,flewelling16}. Middle: GMOS 6100 \AA\ continuum image. The color bar shows the continuum flux values in logarithmic units of erg\,s$^{-1}$\,cm$^{-2}$\,\AA$^{-1}$. Right: [NII]+H$\alpha$ narrow-band image obtained from the GMOS datacube. The color bar shows the flux values in logarithmic units of erg\,s$^{-1}$\,cm$^{-2}$. The blue dashed lines show the orientation of the galaxy major axis as obtained from the 2MASS K-band image \citep{2mass} and the central crosses mark the position of the nucleus, defined as the location of the peak of the continuum emission.  The green contours overlaid to the [NII]+H$\alpha$ image show the inner radio structure of UGC\,8782 at 1360 MHz  presented by \citet{Kukreti22}. In all images,  North is up and East is to the left. } 
    \label{fig:large}
\end{figure*}

During the fit, we tied the velocity and velocity dispersion ($\sigma$) of emission lines produced from the same species (H$\beta$ and H$\alpha$; [O {\sc iii}] $\lambda$5007 and [O {\sc iii}] $\lambda$4959; [N {\sc ii}]$\lambda$6548 and [N {\sc ii}]$\lambda$6583; [S {\sc ii}]$\lambda$6717 and [S {\sc ii}]$\lambda$6731) for each kinematic component. We also fixed the [O {\sc iii}] $\lambda$5007/[O {\sc iii}]$\lambda$4959 and [N {\sc ii}]$\lambda$6583/[N {\sc ii}]$\lambda$6548 intensity line ratios of each component to their theoretical values of 2.98 and 3.06, \citep{Osterbrock2006}, respectively.

 \section{Results}
 
 \subsection{Spatially Resolved Observations}

In Fig.~\ref{fig:large} we present a large scale {\em giy} image of UGC\,8782 (left panel) from Pan-STARRS data archive \citep{chambers16,flewelling16}, along with a continuum (middle panel) and a [N\,{\sc ii}]+H$\alpha$ narrow band (right panel) image obtained from the GMOS data cube.  The GMOS continuum image is obtained by computing the mean flux in a 100\,\AA\ window centred at 6100\,\AA. The blue dashed line shows the orientation of the galaxy's major axis \citep[PA=$50^\circ$; ][]{2mass}. The continuum distribution is elongated along the northeast--southwest direction, has a smaller PA ($\sim35^\circ$ East of North) than for  the large-scale disk, with a secondary extended structure observed to the west of the nucleus at $\sim$1\farcs3. The [N\,{\sc ii}]+H$\alpha$ narrow band image is obtained by integrating the fluxes within a spectral window from 6500 to 6640\,\AA -- including these lines -- using the cube with the contribution of the stellar population component subtracted. Extended gas emission is observed along the major axis of the galaxy, with a similar distribution as observed in the continuum. However, the strongest emission is observed at a structure located at 1\farcs3 west from the nucleus. 

\begin{figure*}
    \centering
    \includegraphics[width=0.99\textwidth]{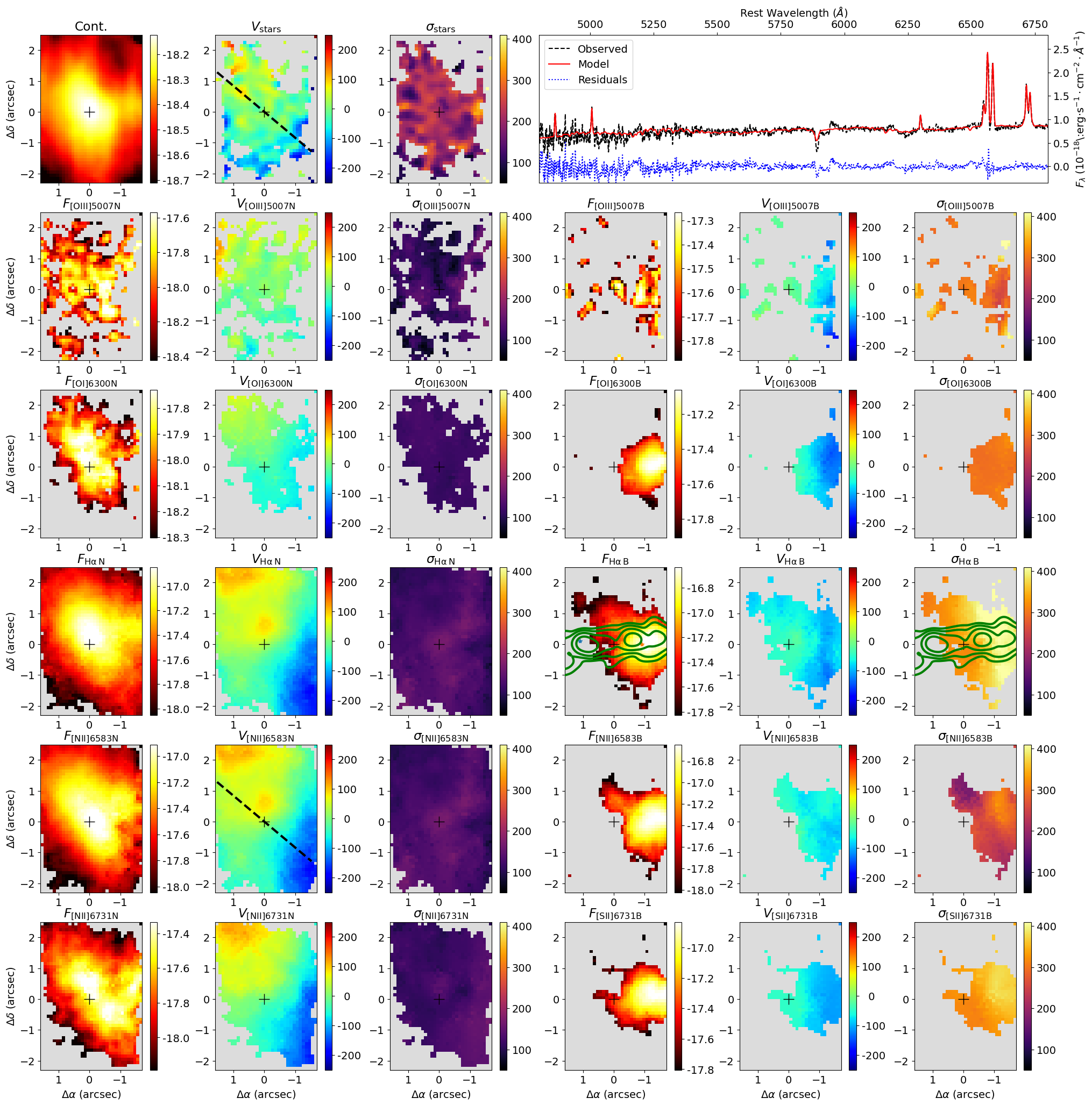} 
\caption{\small Stellar and gas kinematics and distribution in UGC\,8782. The first row shows, from left to right, a continuum image obtained by computing the mean fluxes from the GMOS cube within a 100\,\AA\ window centred at 6000\AA, the stellar velocity field, stellar velocity dispersion map and an example of fitting procedure for the nuclear spaxel. The observed nuclear spectrum  is shown as a black dashed line, the best-fit model is represented in red and the residuals (plus an arbitrary constant) is shown as a dotted line.  The remaining rows show the results for the [O\,{\sc iii}]$\lambda$5007, [O\,{\sc i}]$\lambda$6300, H$\alpha$, [N\,{\sc ii}]$\lambda$6583 and [S\,{\sc ii}]$\lambda$6731 emission lines, from top to bottom. For each emission line, we show, from left to right, the flux distribution, centroid velocity and  velocity dispersion maps for the narrow (first 3 panels) and the broad (last 3 panels) components. The colour bars show the line fluxes in logarithmic units of erg\,s$^{-1}$\,cm$^{-2}$\,spaxel$^{-1}$, the velocity and velocity dispersion maps in km\,s$^{-1}$ and the continuum image in logarithmic units of erg\,s$^{-1}$\,cm$^{-2}$\,\AA$^{-1}\,$spaxel$^{-1}$. The central crosses mark the position of the nucleus, defined as the location of the continuum peak and in all panels the North is up and East to the left. The green contours overlaid in some panels are from the 1360 MHz radio image of UGC\,8782 presented by \citet{Kukreti22}. The black dashed lines show the orientation of the galaxy's major axis \citep[PA=50$^\circ$; ][]{2mass}. Gray regions in the stellar kinematic maps correspond to locations with velocity or $\sigma$ uncertainties larger than 50 km\,s$^{-1}$, while in the emission-line maps these regions correspond to locations where the corresponding emission line density flux is not detected above 3 times the standard deviation of the spectra noise, computed in neighboring spectral regions of each emission line. } 
    \label{fig:maps}
\end{figure*}

Fig.~\ref{fig:maps} presents results for the stellar and gas kinematics and emission distributions. The stellar velocity field presents a velocity gradient consistent with the orientation of the major axis of the large scale disc of PA=50$^\circ$ \citep{2mass} with blueshifts seen to the southwest and redshifts to the northeast of the nucleus. The stellar velocity dispersion map presents  values in the range from 100 to 270 km\,s$^{-1}$ with a mean value of 205km\,s$^{-1}$ and standard deviation of 65 km\,s$^{-1}$. Thus, the observed stellar kinematics in UGC\,8782 is consistent with the behaviour expected for a rotation disc with similar orientation of the large scale disc of the galaxy.

The emission-line flux distributions for the narrow component are elongated along the local stellar disk, following a similar orientation of that of the major axis of the continuum image, for all emission lines, as can be seen in the first column of Fig.~\ref{fig:maps}. The corresponding velocity fields also present a similar velocity behaviour of that in the stellar velocity field, with blueshifts to the southwest and redshifts to the northeast. The gas velocity dispersion maps for the narrow component for all emission lines present values smaller than those observed for the stars, with values lower than 150 km\,s$^{-1}$ at most locations. The kinematics and emission structure of the narrow component is consistent with emission of gas in a rotating disc, similarly to that observed for the stars. The sightly larger velocity amplitude and smaller velocity dispersion values observed for the gas as compared to those of the stars can be explained by projection effects if the gas is located in a thin disc and the stars in a thicker distribution. 
 
The broad component emission is extended mostly to the west of the nucleus for all emission lines. The emission peaks at $\sim$1 arcsec west of the nucleus are close to the border of a radio knot observed at 0\farcs7 west. The centroid velocity of the broad component is blueshifted by about 200 km\,s$^{-1}$ and the $\sigma$ maps present values of up to 400 km\,s$^{-1}$. The green contours overlaid to the H$\alpha$ flux distribution and $\sigma$ map for the broad component are from the 1360 MHz radio image obtained with the MERLIN array and presented by \citet{Kukreti22}. The highest fluxes for the broad components are observed co-spatially with the radio structure to the west and the highest velocity dispersion values are seen surrounding the western radio lobe, suggesting that the emission of the broad component is associated to the interaction of the radio jet with the ambient gas.

\begin{figure}
    \centering
    \includegraphics[width=0.48\textwidth]{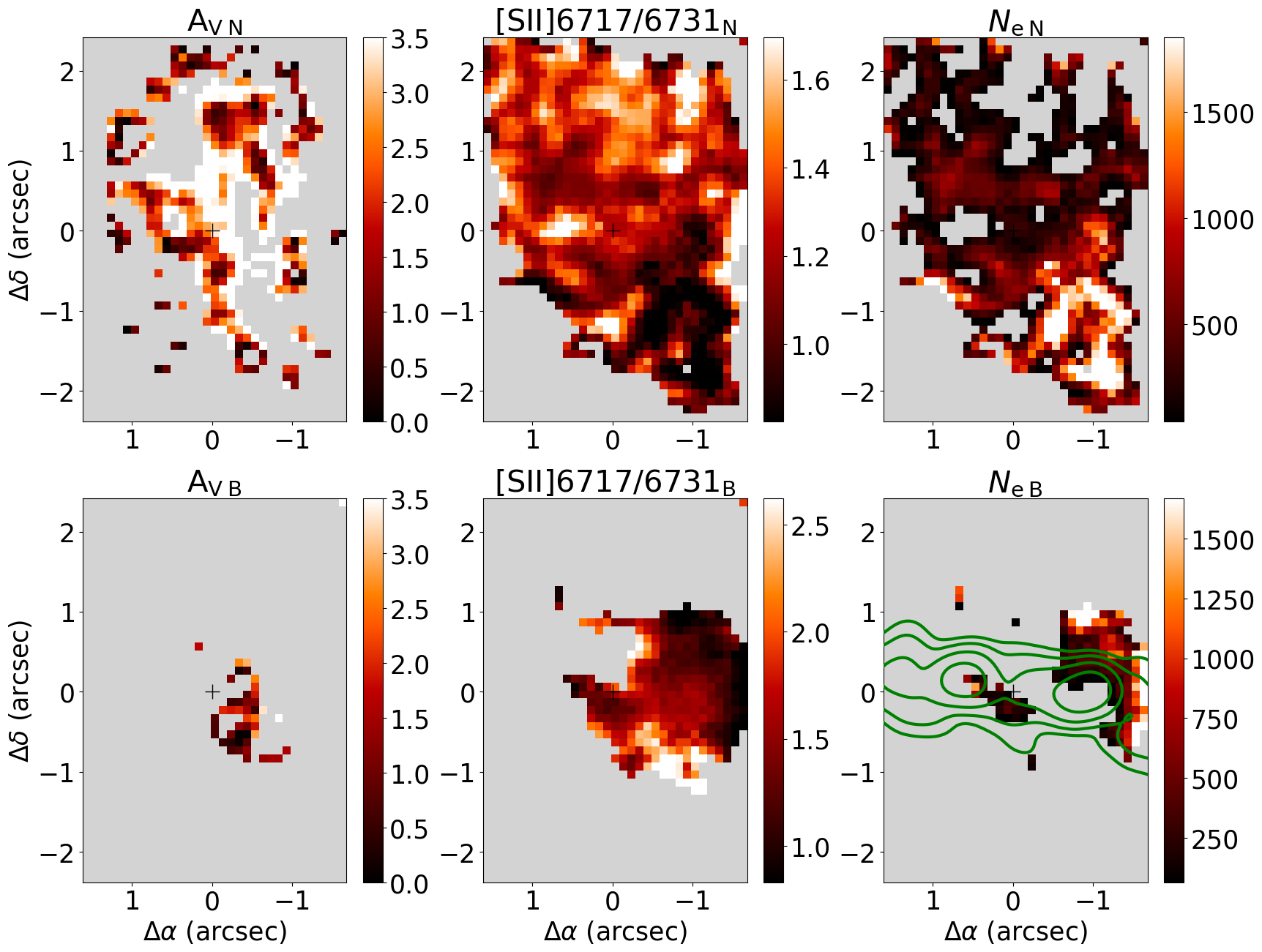} 
\caption{\small Visual extinction (left) in magnitudes, [S\,{\sc ii}]$\lambda$6717/[S\,{\sc ii}]$\lambda$6731 intensity line ratio (middle) and electron density (right) maps for the narrow (top) and broad (bottom) components, in units of cm$^{-3}$. The gray regions correspond to locations where one of both emission lines used in the ratio are not detected above 3-$\sigma$ continuum noise. The green contours overlaid in some panels are from the 1360 MHz radio image of UGC\,8782 presented by \citet{Kukreti22}. } 
    \label{fig:density}
\end{figure}

\begin{figure*}
    \centering
    \includegraphics[width=0.75\textwidth]{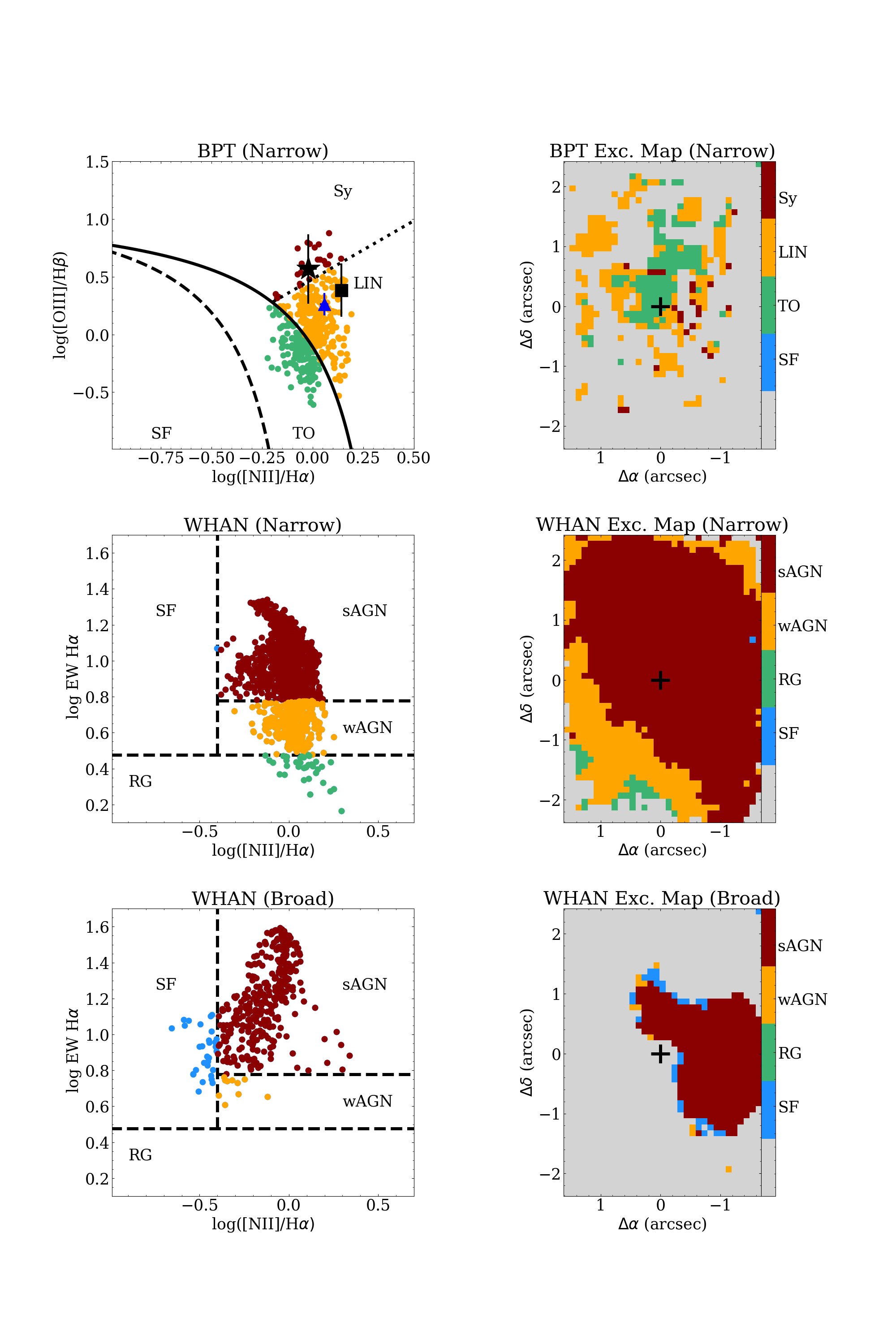} 
 \caption{ Spatially resolved BPT \citep{bpt_1981} and WHAN \citep{Cid_Fernandes_2010,whan_cid_fernandes_2011} diagrams for  UGC\,8782. The top panels show the BPT diagram (left) and the color coded excitation map (right) for the narrow component. 
 The continuous, dashed and dotted lines are the borderlines from \citet{kewley01}, \citet{kauffmann_2003} and  \citet{Cid_Fernandes_2010}, respectively. The blue triangle shows the values of the narrow component for a nuclear aperture of 0\farcs5 radius. The black star and square represent the ratios measured within an aperture of 0\farcs5 centred at 1\farcs0 west of the nucleus (position of the emission peak of the broad component), for the broad and narrow components, respectively. We do not show the BPT diagram for the broad component because the [O\,{\sc iii}] and H$\beta$ lines are detected only in a few spaxels. The middle and bottom panels show the WHAN diagrams and the corresponding excitation maps for the narrow and broad components, respectively. The lines in the WHAN diagram are those defined in \citet{whan_cid_fernandes_2011} and the following nomenclature is used: star-forming galaxies (SF), transition objects (TO), weak AGN (wAGN; i.e. LINERs), strong AGN (sAGN; i.e. Seyferts) and retired galaxies (RG). }    
 \label{fig:bpt}
\end{figure*}

\begin{figure*}
    \centering
    \includegraphics[width=0.85\textwidth]{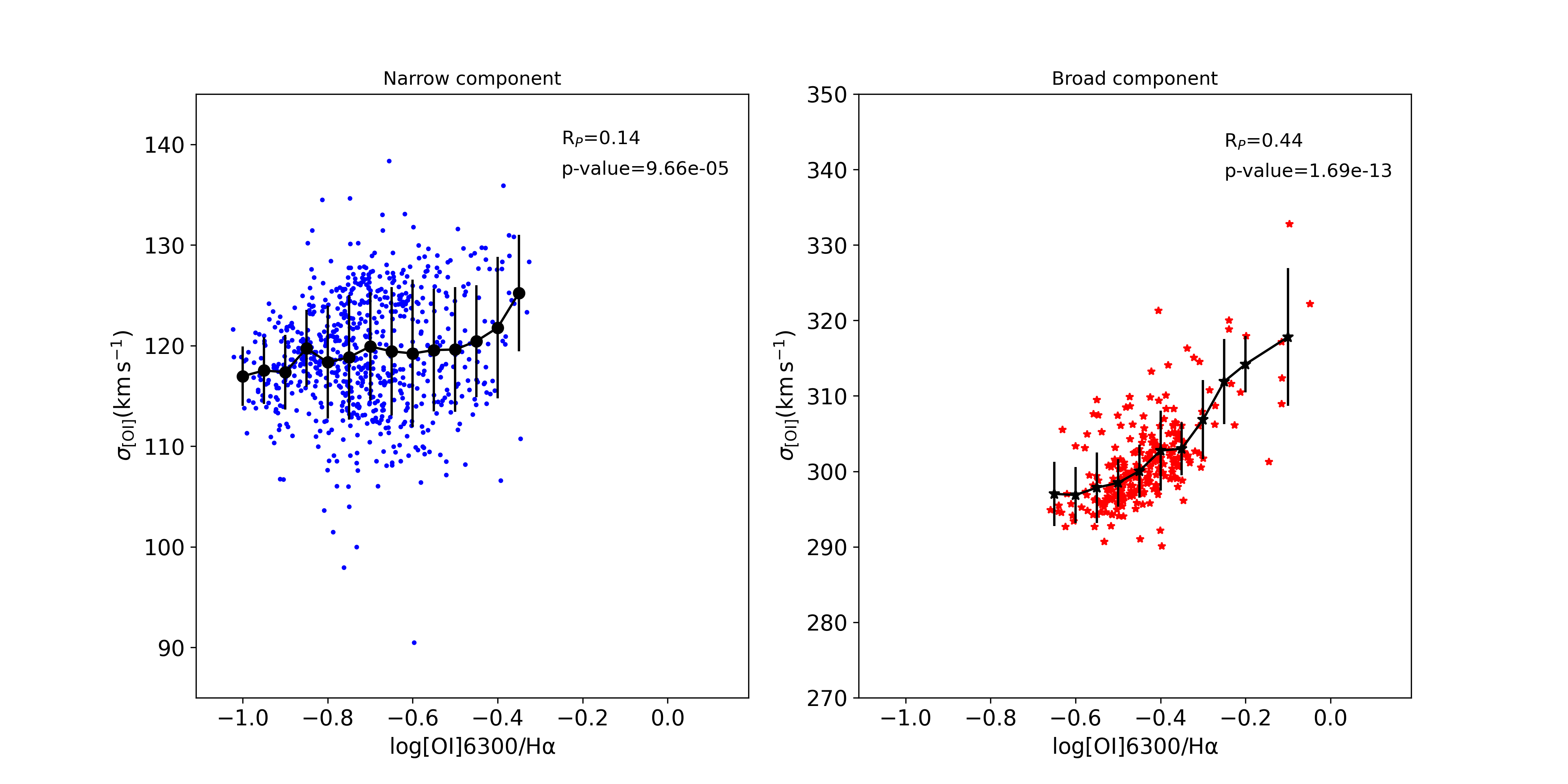} 
\caption{\small Plot of $\sigma_{\rm OI}$  versus  log [O\,{\sc i}]$\lambda$6300/H$\alpha$ for the narrow (left) and broad (right) components. The black dots correspond to mean values computed within bins of 0.05 dex in log [O\,{\sc i}]$\lambda$6300/H$\alpha$. The Pearson correlation coefficient ($R_P$) and $p-value$ are shown in the top-right corner of each panel.}
    \label{fig:OIHA}
\end{figure*}

Following \citet{rogerio_2021}, we estimate the visual extinction (A$_{V}$) using the H$\alpha$/H$\beta$ intensity line ratio, by
\begin{equation}
    A_V = 7.22~\log \left[\frac{(F_{\mathrm{H}\alpha}/F_{\mathrm{H}\beta})_{\rm obs}}{2.86} \right],
\end{equation}
where $(F_{\mathrm{H}\alpha}/F_{\mathrm{H}\beta})_{\rm obs}$ is the observed  H$\alpha$/H$\beta$ flux ratio, adopting the \citet{cardelli_1989} extinction law and an intrinsic flux ratio of $F_{\mathrm{H}\alpha}/F_{\mathrm{H}\beta}=2.86$ for  H {\sc i} case B recombination for an electron temperature of $T_e = 10\,000 $~K and electron density of $N_e = 100~\mbox{cm}^{-3}$ \citep{Osterbrock2006}. The resulting maps for the narrow and broad line components are shown, respectively, in the top and bottom panels of Fig.~\ref{fig:density}.  The A$_{\rm V}$ map for the narrow component shows values in the range from 0 to 3.5 mag, with the highest ones seen at the northwest side of the galaxy, co-spatial with the dust structure seen in the large scale image of Fig.~\ref{fig:large}. The mean value is $\langle {\rm A_V}\rangle$=2.7$\pm$0.1 mag. 
The broad component in the H$\beta$ emission line is detected only in a few spaxels and thus its resulting ${\rm A_V}$ map shows a limited spatial coverage.  The mean value for the broad component is $\langle {\rm A_V}\rangle$=1.7$\pm$0.4 mag.

The central panels of Fig.~\ref{fig:density} show the [S\,{\sc ii}]$\lambda$6717/[S\,{\sc ii}]$\lambda$6731 flux ratio maps for UGC\,8782 for the narrow (top) and broad (bottom) components. For the narrow component, the lowest values of $\sim$ 0.9 are seen to the southwest of the nucleus, while the highest values of up to 1.6 are observed mostly to the northwest and north of the nucleus.  For the broad component, the lowest values of $\sim$ 0.9 are seen to the west of the nucleus, close to the tip of the radio jet, while higher values of up to 2.6 are seen to the south. The [S\,{\sc ii}]$\lambda$6717/[S\,{\sc ii}]$\lambda$6731 ratio can be used to estimate the electron density ($N_e$) for intensity line ratios in the range of $\sim$0.5--1.4 \citep{Osterbrock2006}. Higher  values of [S\,{\sc ii}]$\lambda$6717/[S\,{\sc ii}]$\lambda$6731 are produced in lower density gas, while lower ratios are associated to higher density clouds. 

To estimate the electron density of the emitting gas for both kinematic components, we used the {\sc PyNeb} python package \citep{PyNeb2015} and assume an electron temperature of 1.5$\times10^{4}$ K, a typical value observed in Seyfert nuclei \citep{revalski18a,revalski18b,revalski21,dors20,rogemar21_te,rogemar21_tflut}. The resulting $N_e$ maps are shown in the right panels of Fig.~\ref{fig:density}. The map for the narrow component presents the highest values ($>1700{\,\rm cm^{-3}}$)  to the southwest of the nucleus and the lowest values ($<200{\,\rm cm^{-3}}$) to the northeast, and a mean value of $\langle {\rm N_e}\rangle$=607$\pm$27 ${\,\rm cm^{-3}}$. Most locations with detection of the broad component in the [S\,{\sc ii}] doublet present ratios larger than 1.4, where the relation between the line ratio and $N_e$ becomes flat \citep{Osterbrock2006}. These regions correspond to very low densities ($\lesssim50{\,\rm cm^{-3}}$). The highest densities ($\sim1600{\,\rm cm^{-3}}$) are found close to the edge of the radio jet to the west of the nucleus and the mean value is $\langle {\rm N_e}\rangle$=740$\pm$72 ${\,\rm cm^{-3}}$.

Emission-line ratio diagrams can be used to identify the dominant gas ionization source in a galaxy. In the optical, the BPT \citep{bpt_1981} and WHAN \citep{Cid_Fernandes_2010,whan_cid_fernandes_2011} diagrams are commonly used for this purpose. In the top panels of Fig.~\ref{fig:bpt} we present the spatially resolved [O\,{\sc iii}]$\lambda$5007/H$\beta$ vs.  [N\,{\sc ii}]$\lambda$6583/H$\alpha$ BPT diagram using the fluxes of the narrow line component, as well as a colour coded excitation map (right). The gray regions in the excitation map correspond to locations where at least one line is not detected above the 3-$\sigma$ threshold. Most spaxels with detected emission in all four lines show ratios in the regions occupied by Low-Ionization Nuclear Emission Regions (LINERs) and transition objects (TOs). The blue triangle shown in the BPT diagram corresponds to the nucleus, as obtained using the line fluxes computed within a circular aperture of 0\farcs5 radius.  
The broad component of [O\,{\sc iii}]$\lambda$5007 and H$\beta$ is detected only in a few spaxels around the position of its emission peak (1\farcs0 west of the nucleus) and thus we do not present its BPT diagram, but we compute the integrated fluxes within a 0\farcs5 radius aperture centred at 1\farcs0 west of the nucleus.  The ratios for the narrow and broad component integrated within this aperture are shown as a black square and a black star, respectively. 

The WHAN diagrams for the narrow and broad components, are shown in the middle and bottom panels of Fig.~\ref{fig:bpt}, along with their corresponding excitation maps. The WHAN diagram is particularly useful to separate emission of gas ionized by an AGN -- which results in H$\alpha$ equivalent widths, EW$_{H\alpha}>$3\AA\ -- from those that can be attributed to  hot low-mass evolved stars (HOLMES) with EW$_{H\alpha}<$3\AA\  \citep[][]{Cid_Fernandes_2010,whan_cid_fernandes_2011,agostino21}.  The WHAN diagrams for  UGC\,8782 show that most points are located in the regions occupied by AGNs, and emission of gas excited by HOLMES can be neglected.

UGC\,8782 presents [O\,{\sc i}]$\lambda$6300 extended emission over most of the GMOS field of view (FoV). This emission line can be used as a tracer of shocks in neutral gas \citep[e.g., ][]{allen08,ho14,rogemar20_spitzer}. In Fig.~\ref{fig:OIHA} we present plots of the [O\,{\sc i}] velocity dispersion ($\sigma_{OI}$) versus the [O\,{\sc i}]$\lambda$6300/H$\alpha$ flux line ratio using measurements for the narrow (left panel) and broad (right panel) components. The narrow component presents $100\,{\rm km\,s^{-1}}\lesssim\sigma_{OI}\lesssim130\,{\rm km\,s^{-1}}$ and  $-1.0\lesssim{\rm log(}$[O\,{\sc i}]/H$\alpha\lesssim-0.3$, while for the broad component the ranges of values are $290\,{\rm km\,s^{-1}}\lesssim\sigma_{OI}\lesssim330\,{\rm km\,s^{-1}}$ and  $-0.7\lesssim{\rm log(}$[O\,{\sc i}]/ H$\alpha\lesssim0.0$.  The black points correspond to mean values computed within bins of 0.05 dex in  log [O\,{\sc i}]$\lambda$6300/H$\alpha$, and the error bars represent the standard deviation of the velocity dispersion within each bin.

 \subsection{Integrated spectra at selected positions}

\begin{figure*}
    \centering
    \includegraphics[width=0.99\textwidth]{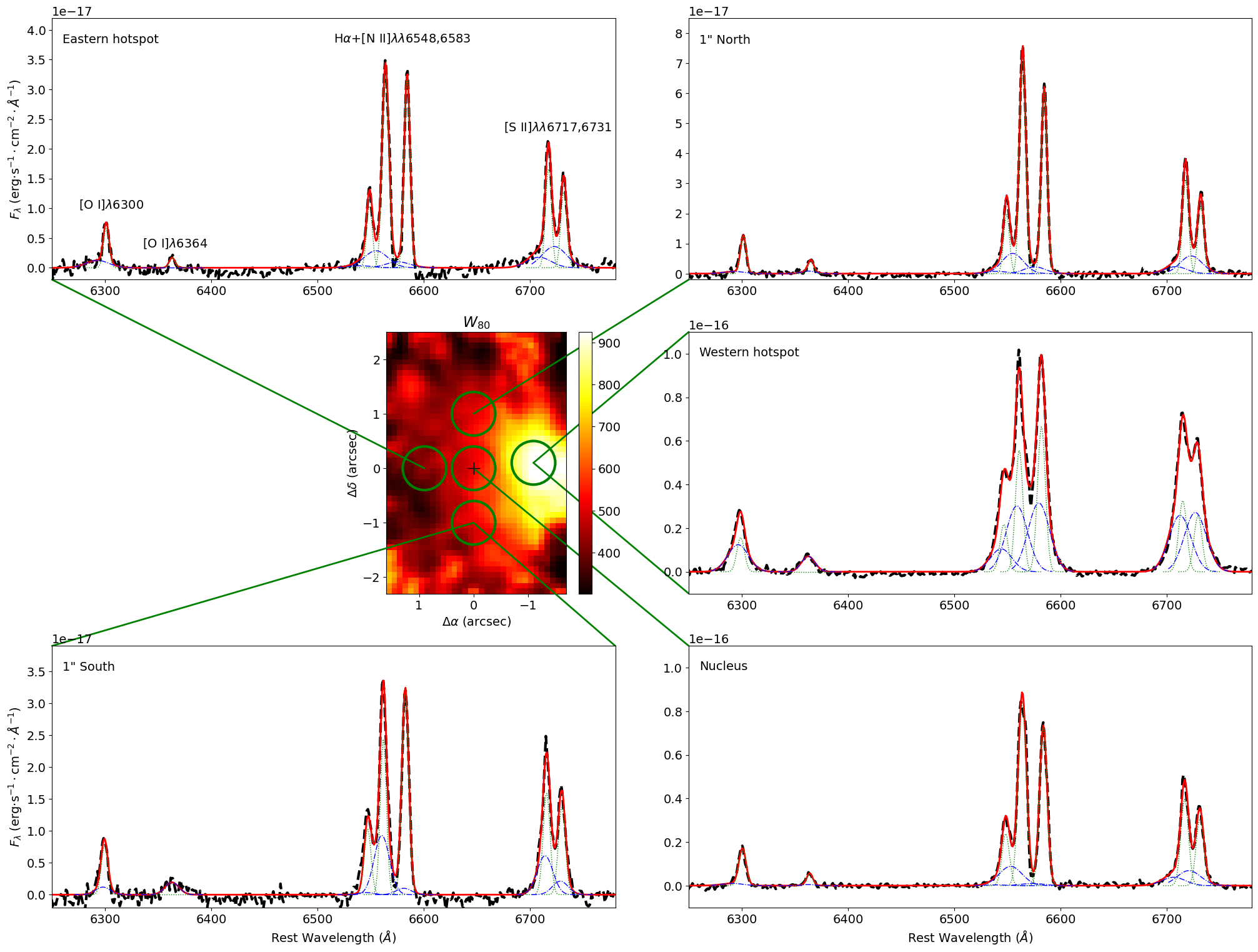} 
        \caption{\small Fits of the emission lines in the red part of the spectra at five positions using integrated spectra extracted within  0\farcs4 radii apertures, as represented by the green circles. The observed spectra with the stellar population contribution subtracted are shown in black and the best-fit models in red. The individual components are shown as green dotted and blue dashed-doted lines for the narrow and broad components, respectively. The central panel shows the $W_{\rm 80}$ map for the H$\alpha$ emission line, with the colour bar representing the values in km\,s$^{-1}$.}  
    \label{fig:2gfit}
\end{figure*}

Previous observations of UGC\,8782 revealed broad emission line components associated to the eastern radio jet, with higher velocities and much fainter emission, as compared to the broad components seen to the west of the nucleus \citep{mahony16}. These authors interpreted this result as the broad components originating from a bipolar outflow, driven by the radio jet. As shown in the previous section, we do not detect this eastern broad component above 3-$\sigma$ noise level in individual spaxels, but there is some extended emission to the north of the nucleus as observed for H$\alpha$ and [N\,{\sc ii}]$\lambda$6583 emission lines (Fig.~\ref{fig:maps}). In order to investigate the origin of the broad component, we have extracted individual spectra within circular apertures of 0\farcs4 radius (comparable to the seeing of the observations) centred at four positions: A) the nucleus; B) the location of the western radio hotspot ($\Delta\alpha=-$1\farcs1; $\Delta\delta$=0\farcs1);  C) the location of the eastern radio hotspot ($\Delta\alpha=$0\farcs9; $\Delta\delta$=0\farcs0); D) 1\farcs0 north of the nucleus and E) 1\farcs0 south of the nucleus. The resulting spectra are shown as black dashed lines in Fig.~\ref{fig:2gfit} and reveal the presence of broad line components in all locations, not only along the radio jet. We followed the same procedure described above and fitted each emission line by two Gaussian curves using the {\sc ifscube} package, keeping the kinematics (velocity and $\sigma$) of the narrow and broad components tied, separately. The resulting models are shown in red in Fig.~\ref{fig:2gfit} and the individual components are shown as green dotted and blue dashed-dotted lines, respectively. 

As in \citet{mahony16} -- who used optical integral-field spectrograph (OASIS) observations of UGC\,8782  with the William Herschel Telescope (WHT) -- we also find that the broad component emission in the western radio hotspot position is about one order of magnitude brighter than that in the eastern hotspot region. In addition, our GMOS data reveals broad component emission in regions perpendicular to the radio jet, as seen in the spectra extracted in regions to the north and to the south of the nucleus in Fig.~\ref{fig:2gfit}. The broad component emission in these regions is stronger than those in the eastern radio hotspot location by a factor of 2, for example, when H$\alpha$ line is considered. The broad component is blueshifted in all locations with velocities of  $-450\pm48$ km\,s$^{-1}$ for the nucleus, $-200\pm57$ km\,s$^{-1}$ (for the eastern hotspot), $-175\pm30$ km\,s$^{-1}$ (western hotspot),   $-140\pm38$ km\,s$^{-1}$ (1\farcs0 south) and  $-365\pm$40 km\,s$^{-1}$ (1\farcs0 north). The widths of the broad component are in the range $380-560$ km\,s$^{-1}$, with the lowest value observed at the south of the nucleus and the highest at the eastern radio hotspot.

In addition, in Fig.~\ref{fig:2gfit} we present the $W_{\rm 80}$ map for the H$\alpha$ emission line. This parameter corresponds to the width of the line profile containing 80 per cent of its total flux and can be used to identify gas outflows which typically present $W_{\rm 80}>500$\,km\,s$^{-1}$ observed in the [O\,{\sc iii]}$\lambda$5007 emission line \citep[e.g.][]{zakamska14,wylezalek20,ruschel-dutra21}. The $W_{\rm 80}$ map for UGC\,8782 presents values larger than this threshold at several location surrounding the nucleus, with the highest values of up to 900 km\,s$^{-1}$ seen to the west of the nucleus along the radio jet.

\section{Discussion}
 
\citet{mahony16} used optical integral field spectroscopy, obtained with the OASIS-IFU on the 4.2 meters William Hershel Telescope,  to study the gas kinematics in the inner 12$\times$9~arcsec$^2$ of UGC\,8782. Their observations covered the [N\,{\sc ii}]+H$\alpha$ and the [S\,{\sc ii}] doublet emission lines with angular sampling ranging from 0\farcs3$\times$0\farcs3 in the central regions to 3\farcs6$\times$2\farcs7 in the outer regions. They detected a blueshifted broad component within the inner $\approx$ 2 arcsec along the east-west direction spanning the region between the inner radio lobes, slightly extended to the north of the nucleus. In addition, these authors analyse long-slit spectra along the positions angles 225$^\circ$ and 93$^\circ$, previously published by \citet{emonts05}, which show that the broad component is detected to distances of up to 5 kpc from the nucleus to both sides, along the radio jet.  Although \citet{mahony16} do not present emission-line flux maps, they mention in their work that the broad component presents much higher peak fluxes in regions to the west of the nucleus, as compared to regions to the east of it.  These authors interpret the broad component as being due to fast outflows driven by the radio jet, in an orientation such that the eastern side is approaching and the western side is receding. The 
 eastern radio jet intercepts the disc, pushing the gas out and the radio jet creates a cocoon that disturbs the gas in all directions \citep{mahony16}.

 Here, we use higher quality  IFU data to further investigate the gas emission structure and kinematics. As in previous works, our GMOS data clearly reveal the presence of a narrow and a broad component, associated to gas emission from the disc and from the outflow. In the following, we will characterize both emission structures.

\subsection{Origin of the gas emission}

The nuclear emission of UGC\,8782 was previously classified as a LINER based on optical line ratios, and interpreted as being produced by shock excitation \citep{emonts05,mahony16}. Our IFU data allow us to further investigate the origin of the gas emission, in a spatially resolved manner. The BPT diagram (Fig.~\ref{fig:bpt}) shows line ratios mostly in the region occupied by LINERs and transition objects, with the nuclear emission being consistent with a LINER-type ionization. At the location of the western radio hotspot, the line ratios for the narrow component are also consistent with a LINER-type ionization and for the broad component the observed line ratios are close to the borderline separating Seyferts and LINERs in the WHAN diagram, as defined in \citet{Cid_Fernandes_2010}.  These results are consistent with the classification of UGC\,8782 as a host of an AGN, based on the plot of the ${\rm PAH 6.2\,\mu m}$ equivalent width (EW$_{\rm PAH 6.2\,\mu m}$) versus the ${\rm H_2 S(3)/PAH 11.3\,\mu m}$ intensity line ratio using Spitzer data \citep[EW$_{\rm PAH 6.2\,\mu m}\approx0.09\,\mu$m  and ${\rm H_2 S(3)/PAH 11.3\,\mu m}\approx0.15$ for UGC\,8782;][]{lambrides19}.

Although the BPT and WHAN diagnostic diagrams are useful to separate photoionized gas by star forming regions, AGN and HOLMES, they are not sensitive to possible contributions from shocks \citep{dors21,agostino21}. The [O\,{\sc i}]$\lambda$6300/H$\alpha$ intensity line ratio is a well known tracer of shocks \citep{monreal-ibero06,monreal-ibero10,rich11,rich15}. Shocks may be the dominant excitation mechanism of the [O\,{\sc i}] emission if its velocity dispersion is larger than $\sim$150 km\,s$^{-1}$ and log\,[O\,{\sc i}]$\lambda6300$/H$\alpha \gtrsim -1.0$ \citep[e.g.][]{ho14}. The plots of $\sigma_{\rm OI}$  versus  log [O\,{\sc i}]$\lambda$6300/H$\alpha$ shown in Fig.~\ref{fig:OIHA} reveal that, although log\,[O\,{\sc i}]$\lambda6300$/H$\alpha > -1.0$ for the disc component, the velocity dispersion is low (100--130 km$\,$s$^{-1}$), suggesting that photoionization may be the be main mechanism in action. However, shocks with these velocities may also be able to excite the gas \citep{ho14}. Indeed, an additional contribution of shocks cannot be discarded as a weak correlation (Pearson correlation coefficient $R_P$=0.14) is observed between [O\,{\sc i}]$\lambda6300$/H$\alpha$ and $\sigma_{\rm OI}$ for the narrow component. The highest values of both parameters are observed in regions along the disc, outside the AGN ionization cone. This result is consistent with the detection of shock ionization by AGN winds in regions outside the AGN ionization structure, where shocked gas emission can be easily observed \citep{rogemar21_te}, and with the detection of a broad component in the integrated spectra shown in  Fig.~\ref{fig:2gfit}.  On the other hand, the broad outflow component shows much higher [O\,{\sc i}]$\lambda6300$/H$\alpha$ and $\sigma_{\rm OI}$ and there is a stronger correlation between these two parameters ($R_P$=0.44), which suggests that shock excitation is important in the production of emission of the outflowing gas. 

Thus, the main excitation mechanism of the gas in the disc may be photoionization by the central AGN of UGC\,8782 as indicated by the observed emission line ratios and kinematics, with an additional contribution of shocks. On the other hand, the enhanced gas emission associated to radio hotspots, the observed intensity line ratios and the correlation between [O\,{\sc i}]$\lambda6300$/H$\alpha$ and $\sigma_{\rm OI}$ indicate that shocks play an important role in the excitation of the gas in the outflow.

\subsection{The disc component}

\begin{figure*}
    \centering
    \includegraphics[width=0.99\textwidth]{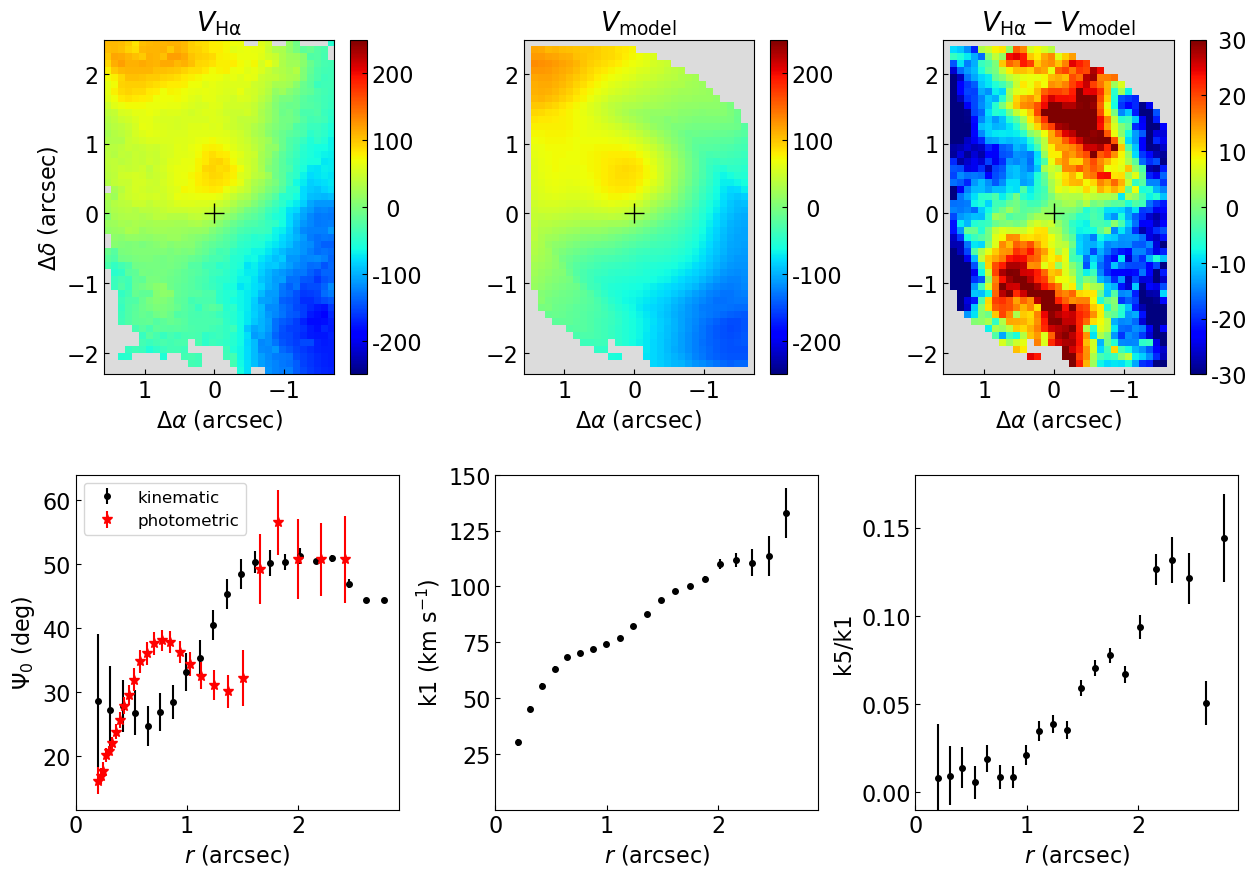} 
        \caption{\small  Top panels, from left to right: observed H$\alpha$ velocity field for the narrow component, best-fit model obtained with the {\sc kinemetry} code and residual velocity map, obtained by subtracting the model from the observed velocities.
        Bottom panels: the left panel shows plot of the position angle (PA) of the major axis of the modelled ellipses along the radius. The black circles represent the kinematic PAs, and the red stars show the photometric PAs obtained by fitting the continuum image using the {\sc photutils} package.   The central  and right panels show the harmonic expansion coefficients $k_1$ and $k_5/k_1$ from the {\sc kinemetry} model. }  
    \label{fig:kinemetry}
\end{figure*}

The gas velocity field for the disc component (Fig.~\ref{fig:maps}) shows a clear rotation pattern for all emission lines, with the line of nodes oriented approximately along the orientation of the major axis of the galaxy's large scale disc \citep[PA=50$^\circ$; ][]{2mass}. This behaviour is similar to that observed for the stars (Fig.~\ref{fig:maps}) and in cold molecular gas, traced by the CO emission \citep{labiano14}. However, deviation from pure rotation is also observed in the ionized gas, as a discontinuity in velocities seen at $\sim$1 arcsec northeast of the nucleus. In order to better constrain the gas kinematics, we use the {\sc kinemetry} method \citep{Krajnovic2006_kinemetry} to model the gas velocity field from the H$\alpha$ narrow component. This method is a generalization of the surface photometry technique and can be used to model the observed velocity field by a set of ellipses under the assumption that the velocity profile along each ellipse satisfies a cosine law, using the Fourier expansion technique \citep{Krajnovic2006_kinemetry}.

Fig.~\ref{fig:kinemetry} shows, in the top panels, the observed H$\alpha$ velocity field for the disc (narrow) component (left panel), the resulting best-fit model obtained by the {\sc kinemetry} method (central  panel) and the residual map obtained by subtracting the model from the observed velocities (right-panel). The bottom-left panel shows the radial variation of the kinematic position angle (PA) of the major axis of the fitted ellipses (in black) and the photometric PA (in red) obtained by modeling the continuum image (Fig.~\ref{fig:maps}) using the {\sc photutils} package \citep{larry_bradley_2020_4044744}. The bottom-central  and bottom-right panels show the harmonic coefficients $k_1$ and $k_5/k_1$, respectively. The $k_1$ term describes the velocity amplitude of bulk gas motions, while the  $k_5/k_1$ term quantifies higher-order deviations from pure rotation \citep{Krajnovic2006_kinemetry}.

The photometric and kinematic PA of the major axis of the ellipses approximately follow each other, with the highest offset of less than 10$^\circ$ observed at $\sim$0\farcs7 and at 1\farcs5 from the nucleus. These discrepancies are observed at the same distances of a redshifted structure north of the nucleus and an excess of blueshifts to the north-west of it, observed in the velocity field. The position angle of the major axis changes from $\Psi_0=\sim20^\circ$ close to the nucleus to $\Psi_0=\sim50^\circ$ in the outer regions, becoming similar to the values derived for the large scale disc using images in different bands \citep[$50-68^\circ$][]{Mackay71,sdss_Dr6,2mass} and from the analysis of the kinematics of the cold gas \citep{labiano14}. The near side of the disc corresponds to the northwestern side of the galaxy, where obscuration is higher \citep{floyd06,labiano14}. A higher extinction at the northwestern side of the galaxy is also confirmed by our GMOS data in a map for the ratio between the red and blue stellar continuum, resulting in higher values to the northwestern side of the disc.

The rotation curve (bottom central panel in Fig.~\ref{fig:kinemetry}) shows increasing velocity values outward from the nucleus, reaching a maximum value of $\sim$130 km\,s$^{-1}$. The  $k_5/k_1$ parameter shows values close to zero in the inner $\sim$1 arcsec, and growing outwards, reaching a maximum value of $\sim$0.15. This indicates that beyond the inner arcsec, the gas kinematics shows deviations that cannot be explained by regular rotation under the galaxy's gravitational potential. The residual map (top-right panel of Fig.~\ref{fig:kinemetry}) shows low values ($\lesssim$30\,km\,s$^{-1}$), but with a systematic pattern, supporting an additional kinematic component.  Redshifted regions are observed approximately along the minor axis of the galaxy, while blueshifted residuals are seen mainly along the major axis of the galaxy.  The complex gas kinematics observed in the disc component of UGC\,8782 is likely produced by a merger event with a gas-rich galaxy, providing a gas reservoir to feed the central SMBH as suggested by previous studies \citep[e.g.][]{martel99,evans99,beswick02,emonts05}.

We can estimate the mass of ionized gas in the disc by 
\begin{equation}\label{ma}
     M_{\rm ion}=1.4 N_{\rm e} m_{\rm p} V f,  
\end{equation}
where $m_{\rm p}$ is the mass of the proton, $N_{\rm e}$ is the electron density,  $V$ is the volume of the emitting gas,  the 1.4 factor is included to account for the contribution of He, and $f$ is the filling factor which can be obtained from
\begin{equation}\label{fa}
L_{\rm H\alpha}\approx {4\pi j_{\rm H\alpha} V f}, 
\end{equation}
where $L_{\rm H\alpha}$ is the H$\alpha$ luminosity and $4\pi\,j_{\rm H\alpha}/N_e^2=3.558\times10^{-25}$ erg\, cm$^{-3}$\,s$^{-1}$ assuming the case B H\,{\sc ii} recombination for the low-density limit and an electron temperature of 10\,000 K \citep{Osterbrock2006}. Replacing $f$ from Eq.~\ref{fa} into Eq.~\ref{ma}, we obtain
\begin{equation}\label{eq:mion}
     M_{\rm ion}\approx\,7\times\frac{L_{\rm H\alpha}}{N_{\rm e}}. 
\end{equation}

We calculate  $L_{\rm H\alpha}$ from the H$\alpha$ fluxes for the disc component in each spaxel, after correcting them for extinction using the $A_{\rm V}$ map from Fig.~\ref{fig:density} and the extinction law of \citet{cardelli_1989}. For spaxels with no $A_{\rm V}$ estimates, we use the mean $A_{\rm V}$ value to correct the H$\alpha$ flux. Using the $N_{\rm e}$ values for the disc component and using their mean value in spaxels with no $N_{\rm e}$ measurements, we obtain $M_{\rm ion}=(1.1\pm0.4)\times10^7$ M$\odot$. The mass of ionized gas in the disc is about three orders of magnitude lower than the mass of cold gas in  UGC\,8782, of 2.2$\times$10$^{10}$\,M$_\odot$  \citep{evans99,labiano14}.

\subsection{Properties of the outflows}

The geometry of the multi-phase gas outflows in UGC\,8782 is an intriguing question. Outflows in neutral gas are co-spatial with the western radio hotspot \citep{mahony13}, while previous studies of the ionized gas kinematics found that the highest velocity component arises from a region co-spatial with the eastern radio hotspot \citep{mahony16}. Shock-heated gas emission is detected in X-rays at the nucleus and inner radio jet, as well as in diffuse X-ray emission to the north and to the south of the nucleus \citep{lanz15}. The cold molecular counterpart of the outflows has not yet been detected, and IRAM Plateau de Bure interferometer observations of the CO emission give an upper limit of $7.1\times10^8$ M$_\odot$ for the mass of the cold gas outflow \citep{labiano14}. The eastern side of the inner radio jet is approaching us making an angle of $\sim$50$^\circ$ with the line of sight \citep{Beswick04}, consistent with the orientation of the galaxy 
such that the northwestern side is nearer to us. The inclination of the disk is $i=50\pm5^\circ$ (relative to the plane of the sky) as obtained from ellipse fitting of the continuum image and consistent with the large scale disc \citep[$i\approx53^\circ$][]{2mass}.

\citet{mahony16} interpreted the ionized gas outflows in UGC\,8782 as being driven by the radio jet, and a cocoon structure surrounding the radio jet disturbing the gas in the central region of the galaxy. They found that the outflows are detected in blueshifts on both sides of the nucleus, with higher velocities observed to the eastern side attributed to the interaction of the approaching (eastern side) radio jet with the ambient gas. As shown in Fig.~\ref{fig:2gfit}, our GMOS data reveal the presence of a blueshifted broad component not only at the position of the radio hotspots but also from regions away of the radio structure. In addition, the velocities of the outflow are similar at locations co-spatial with the eastern and western radio hotspots.  As already noticed by \citet{mahony16}, the flux of the broad component at the eastern hotspot is very low and marginally detected in their data, leading to a high uncertainty in the velocity of the outflow. Indeed, the example of line fits shown by these authors in their Fig.~3 seems to  slightly overestimate the blue wing of the [S\,{\sc ii}] lines leading to a higher inferred velocity, as compared to the most prominent outflow component detected at the location of the western radio hotspot.

In agreement with previous observations, our data also support a contribution of shock excitation of the gas in the outflow, as discussed above.  The close association between the enhanced emission structures seen in ionized gas with the radio structures indicates that the shocks are produced by jet-cloud interaction. From the spectra extracted on selected locations (Fig.~\ref{fig:2gfit}) we find that the outflow component is blueshifted at all positions, with the highest velocity observed for the nucleus. The outflows are seen in blueshifts at all locations surrounding the nucleus if they are produced by the cocoon created by the radio jet \citep{mahony16}, with the enhanced emission to the western side of the galaxy being produced by the interaction of the radio jet with the ambient gas of the galaxy. Enhanced emission-line widths in regions perpendicularly to the radio jet have been observed in nearby Seyfert galaxies, but with centroid velocities close to the systemic velocity of the galaxies, and interpreted as being due to jets perturbing the gas in the disc \citep[e.g.][]{venturi21,Girdhar22}, equatorial outflows predicted by torus wind models \citep[e.g.][]{Elitzur12,honig13,rogemar14_n5929_let} or due to much wider opening angle outflows as compared to the AGN ionization cones  \citep[e.g.][]{rogemar21_te}. The observed blueshifted velocities at all locations around the nucleus favor a scenario in which the ionized outflows in UGC\,8782 present a spherical geometry, consistent with some theoretical predictions \citep{wagner12,Ishibashi19}.

To compute the mass of ionized gas in the outflow, first we correct the flux of the H$\alpha$ broad component at each spaxel using the $A_V$ values (Fig.~\ref{fig:density}) for the broad component and the extinction law of \citet{cardelli_1989}. For spaxels with no $A_V$ estimations, we use its mean value. Then, we compute the mass of the gas in each spaxel using Eq.~\ref{eq:mion} and the observed $N_e$ values for the outflow from Fig.~\ref{fig:density} or its mean value for spaxels with no density measurements. Finally, we sum the mass contributions of all spaxels resulting in a mass of ionized gas in the outflow of $M_{\rm out}=(3.3\pm1.1)\times10^{6}$\,M$_\odot$. This value corresponds to about 25 per cent of the total mass of ionized gas in the central region of UGC\,8782.

\begin{figure}
    \centering
    \includegraphics[width=0.48\textwidth]{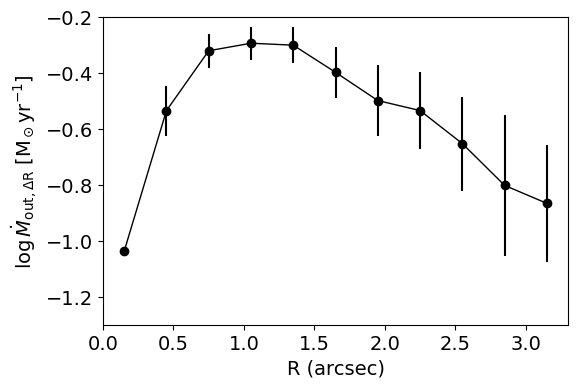} 
        \includegraphics[width=0.48\textwidth]{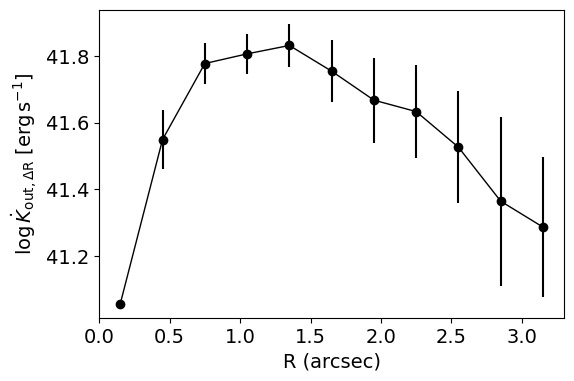} 
        \caption{\small  Radial profiles of the mass-outflow rate (top panel) and kinetic power (bottom panel) of the ionised outflows in  UGC\,8782. }  
    \label{fig:radial}
\end{figure}

We estimate the mass-outflow rate and kinetic power of the ionized gas outflows within circular rings with width $\Delta R=$0\farcs3 centred at the nucleus by 
    \begin{equation}
         \dot{M}_{\rm out,\Delta R}=\sum_i\frac{M^i_{\rm out,\Delta R} V^i_{\rm out,\Delta R}}{\Delta R},
     \end{equation}
     and
    \begin{equation}
         \dot{K}_{\rm out,\Delta R}=\sum_i\frac{M^i_{\rm out,\Delta R} (V^i_{\rm out,\Delta R})^3}{2\Delta R},
     \end{equation}
where the sum is performed over velocity bins corresponding to the spectral sampling of the final data cube ($\sim$45\,km\,s$^{-1}$) with velocities in the range from $-3000$ to $3000$ km\,s$^{-1}$, $M^i_{\rm out,\Delta R}$  is the mass of ionized gas within the ring at the velocity channel $i$  and $V^i_{\rm out,\Delta R}$ corresponds to the absolute value of the velocity of the outflow.  We constructed velocity channel maps for the H$\alpha$ outflow component, using a datacube constructed by subtracting the models of the stellar population contribution and of the narrow H$\alpha$ component and the [N\,{\sc ii}] emission lines. To compute the mass within each channel and radial bin, we adopt the mean electron density value for the outflow component.  Fig.~\ref{fig:radial} presents the resulting radial profiles of mass-outflow rate (top panel) and kinetic power of the outflows (bottom panel). The error bars corresponds to the uncertainties propagated from the flux, velocity and electron density. The peaks of $ \dot{M}_{\rm out,\Delta R}$ and $\dot{K}_{\rm out,\Delta R}$ occur at $\sim$1 arcsec, as expected by the enhanced emission at the location of the western radio hotspot, reaching values of $0.5\pm0.1$ M$_\odot$\,yr$^{-1}$ and (6.8$\pm$1.1)$\times$10$^{41}$ erg\,s$^{-1}$, respectively. These are the lower limits because where there is ionized gas, there must be neutral gas as well because otherwise the clouds get over-ionized to a much higher ionization level than observed \citep{dempsey18}. The peak mass outflow rate derived here is about 3 times larger than the value derived by \citet{mahony16}. This discrepancy is due to the different assumptions by these authors and in this work, and specifically due to the correction of the H$\alpha$ fluxes for extinction, not performed in the previous work.  
The kinetic power of the outflows is about two orders of magnitude smaller than the kinetic power of the radio jets in UGC\,8782, of $2-4\times$10$^{43}$ erg\,s$^{-1}$ \citep{lanz15}, consistent with a jet-driven outflow. 

We estimate the AGN bolometric luminosity ($L_{\rm bol}$) from the [O\,{\sc iii}]$\lambda$5007 luminosity by $L_{\rm bol}=3500 L_{\rm [O\,III]}$ \citep{Heckman04}. Using the integrated  [O\,{\sc iii}]$\lambda$5007 luminosity, corrected by dust extinction the extinction law of \citet{cardelli_1989}, we obtain $L_{\rm bol}=(2.2\pm0.7)\times10^{43}$ erg\,s$^{-1}$. Using the relation of $L_{\rm bol}$ with the H$\alpha$ luminosity, $L_{\rm bol}=2000 L_{\rm H\alpha}$ \citep{Sikora13} the resulting bolometric luminosity is  $L_{\rm bol}=(2.5\pm0.4)\times10^{43}$ erg\,s$^{-1}$.
Finally, using the hard X-ray bolometric correction from \citet{netzer19} and the observed 2--10 keV flux from \citet{ueda05}, we obtain $L_{\rm bol}\approx7.2\times10^{43}$ erg\,s$^{-1}$. An advantage of using X-ray observations is that it is less sensitive to dust obscuration and to scattered or reprocessed emission than optical wavelengths, being a direct tracer of the AGN emission. 

The kinetic power of the ionized outflows corresponds to $\sim$1--3 per cent of the bolometric luminosity of the AGN in UGC\,8782, depending on the tracer used to estimate the bolometric luminosity.  This indicates that the ionized gas outflows could be powerful enough to suppress the star formation in the host galaxy as the  kinetic coupling efficiency of the outflow is larger than the minimum value required by simulations for AGN feedback to have an effective impact on the host galaxies \citep[e.g.][]{dimatteo05,hopkins_elvis10,dubois_horizon_14,Schaye_eagle_15,weinberger17,Harrison_2018}. In addition, the kinetic energy of the outflows represents less than 20 per cent of the total energy released by the outflow and the contribution from the other gas phases is expected to be non negligible. \citep{richings18b}. On the other hand, it is worth mentioning that the uncertainties in the outflow properties are usually high (up to two orders of magnitude), due to assumptions on their densities and geometry \citep{Davies2020,revalski22,molina22,rogemar_2023_extended}.

\section{Conclusions}

We used GMOS-IFU observations of the radio loud AGN host UGC\,8782 at a spatial resolution of $\sim$725 pc to study the gas emission structure and kinematics in the inner 3.4$\times$4.9 kpc$^2$ of the galaxy. Our main conclusions are:

\begin{itemize}

    \item The emitting gas presents high velocity dispersion as quantified by the $W_{\rm 80}$ maps, with values larger than 500\,km\,s$^{-1}$ at several locations. The highest values are observed west ($\sim 1$\,kpc) of the nucleus in regions co-spatial with the inner radio jet, indicating a jet-cloud interaction. 
    \item The profiles of each emission line are fitted by two Gaussian curves: a narrow ($\sigma\lesssim200$ km\,s$^{-1}$) and a broad ($\sigma\gtrsim200$ km\,s$^{-1}$) component. The narrow component traces the emission of the gas in the galaxy disc, while the broad component is consistent with gas outflows.
    \item The narrow component gas kinematics shows a pattern of disc rotation, similar to that of the stars. However, differences between the  observed H$\alpha$ velocity field and a rotation disc model of up to 30 km\,s$^{-1}$ are also present,  likely associated to disturbed gas kinematics due to a previous galaxy merger process.
    \item The broad component is blueshifted by $\sim150-500$\,km\,s$^{-1}$ relative to the systemic velocity of the galaxy and has a velocity dispersion of up to $\sigma\approx400$\,km\,s$^{-1}$. This component is detected for individual spaxels mainly at locations to the west and north of the nucleus, but it is observed in integrated spectra at several locations surrounding the nucleus. 
    \item The BPT and WHAN diagnostic diagrams show that both disc and outflow components present values in the AGN region. Shock gas excitation is also supported for the outflow component by the enhanced emission co-spatial with the western radio hotspot and by a correlation between [O\,{\sc i}]$\lambda6300$/H$\alpha$ and $\sigma_{\rm OI}$, known shock tracers.
    \item  We estimate the mass outflow rate and kinetic power of the ionized gas outflows at different distances from the nucleus. The highest values of  these properties are observed at $\sim1$\,kpc from the nucleus with values of $\dot{M}_{\rm out,\Delta R}=0.5\pm0.1$ M$_\odot$\,yr$^{-1}$ and $\dot{K}_{\rm out,\Delta R} =$(6.8$\pm$1.1)$\times$10$^{41}$ erg\,s$^{-1}$. The kinetic power of the outflows corresponds to 1--3 per cent of the AGN bolometric luminosity. 
    
\end{itemize}

\section*{Acknowledgements}
We thank to an anonymous referee for their constructive comments that helped improve this manuscript.
Based on observations obtained at the Gemini Observatory, which is operated by the Association of Universities for Research in Astronomy, Inc., under a cooperative agreement with the NSF on behalf of the Gemini partnership: the National Science Foundation (United States), National Research Council (Canada), CONICYT (Chile), Ministerio de Ciencia, Tecnolog\'{i}a e Innovaci\'{o}n Productiva (Argentina), Minist\'{e}rio da Ci\^{e}ncia, Tecnologia e Inova\c{c}\~{a}o (Brazil), and Korea Astronomy and Space Science Institute (Republic of Korea). 
This research has made use of NASA's Astrophysics Data System Bibliographic Services. This research has made use of the NASA/IPAC Extragalactic Database (NED), which is operated by the Jet Propulsion Laboratory, California Institute of Technology, under contract with the National Aeronautics and Space Administration.
 R.A.R. acknowledges the support from Conselho Nacional de Desenvolvimento Cient\'ifico e Tecnol\'ogico (CNPq)  and Funda\c c\~ao de Amparo \`a pesquisa do Estado do Rio Grande do Sul (FAPERGS).
 R.R. thanks to CNPq (Proj. 311223/2020-6,  304927/2017-1 and  400352/2016-8), FAPERGS (Proj. 16/2551-0000251-7 and 19/1750-2). M.B. and G.L.O thank to Coordena\c{c}\~ao de Aperfei\c{c}oamento de Pessoal de N\'{i}vel Superior (CAPES, Proj. 0001).

\section*{Data Availability}
The processed data used in this article will be shared on reasonable request to the corresponding author.




\bibliographystyle{mnras}
\bibliography{paper_r1} 





\bsp	
\label{lastpage}
\end{document}